\documentclass[
  ,draft            
  ]
  {aipproc}

\layoutstyle{6x9}

\usepackage{amsmath}
\usepackage{amssymb}
\usepackage{graphicx}
\usepackage{pst-all,pstricks-add}

\newcommand{\ket}[1]{\left| #1 \right>}
\newcommand{\bra}[1]{\left< #1 \right|}

\newcommand{\innerprod}[2]{\left< #1 \vert #2 \right>}
\newcommand{\commut}[2]{\bigl[ #1, #2 \bigr]}

\newcommand{\rhoi}{\hat{\rho}_\mathrm{i}}
\newcommand{\rhof}{\hat{\rho}_\mathrm{f}}
\newcommand{\rhofconst}{\hat{\rho}_{\mathrm{f} \, \mathrm{const}}}
\newcommand{\rhofbal}{\hat{\rho}_{\mathrm{f} \, \mathrm{bal}}}
\newcommand{\Econst}{\hat{E}_\mathrm{const}}
\newcommand{\Ebal}{\hat{E}_\mathrm{bal}}

\DeclareMathOperator{\Trace}{\mathrm{Tr}}
\DeclareMathOperator{\Prob}{\mathrm{Pr}}

\begin{document}

\title{Discrimination of Unitary Transformations and Quantum Algorithms}


\author{David Collins}{
  address={Department of Physical and Environmental Sciences, Mesa State College, Grand Junction, Colorado, USA},
  ,email={dacollin@mesastate.edu}
}

\begin{abstract}
   Quantum algorithms are typically understood in terms of the evolution of a multi-qubit quantum system under a prescribed sequence  of unitary transformations. The input to the algorithm prescribes some of the unitary transformations in the sequence with others remaining fixed. For oracle query algorithms, the input determines the oracle unitary transformation. Such algorithms can be regarded as devices for discriminating amongst a set of unitary transformations. The question arises: ``Given a set of known oracle unitary transformations, to what extent is it possible to discriminate amongst them?'' We investigate this for the Deutsch-Jozsa problem. The task of discriminating amongst the admissible oracle unitary transformations results in an exhaustive collection of algorithms which can solve the problem with certainty.
\end{abstract}

\maketitle



\section{Oracle algorithms and discrimination of quantum operations}

Quantum algorithms~\cite{deutsch92,simon94,shor97,grover97} are usually solve computational problems with the aid of a binary oracle function $f: \{0,1\}^n \mapsto \{0,1\}^m,$ which depends on the problem. The simplest case, where $m=1,$ occurs in the Deutsch-Jozsa~\cite{deutsch92} and Grover's~\cite{grover97} algorithms. In quantum algorithms a given oracle, $f$, is invoked via unitary transformation, $\hat{U}_f,$ whose structure depends on the nature of the oracle. For example, in both the Deutsch-Jozsa algorithm~\cite{collins98} and Grover's algorithm the oracle is defined on computational basis states, $\ket{x} \equiv \ket{x_n\ldots x_1}$ with $x_i\in \left\{0,1\right\},$ to be $\hat{U}_f \ket{x} = \left( -1\right)^{f(x)}\, \ket{x}$ and this is extended linearly to superpositions of computational basis states. The general structure of such oracle algorithms is encapsulated in an algorithm unitary
\begin{equation}
 \hat{U}_\mathrm{alg} =  \hat{V}_M \hat{U}_f \ldots \hat{U}_f \hat{V}_1 \hat{U}_f \hat{V}_0
\end{equation} 
where $\hat{V}_0, \ldots, \hat{V}_M$ are oracle independent unitary transformations and the oracle is invoked $M$ times. This is applied to a quantum system in an oracle independent initial state $\ket{\Psi_i}$, giving an oracle dependent final state $\ket{\Psi_f} = \hat{U}_\mathrm{alg} \ket{\Psi_i},$ upon which a computational basis measurement is performed. It is important to note that \emph{the input to the algorithm is the oracle unitary} and not the initial state. The output from the algorithm potentially identifies the oracle unitary or a class of oracle unitaries. For example, in Grover's algorithm for searching a database one marked item at location $s,$ $f(x) = 0$ whenever $x \neq s$ and $f(s)=1.$ The standard Grover's algorithm terminates in a computational basis measurement which yields $s$ with high probability. Since the admissible oracles  for this problem can be labeled by $s$, the algorithm \emph{identifies the input oracle unitary} with high probability. We thus regard the quantum algorithm as a tool for discriminating between classes of admissible input oracle unitaries.

The problem of discriminating between unitary transformations is usually reduced to a quantum state discrimination problem~\cite{acin01,sacchi05}. This considers application of one unitary from a collection of known possible unitaries, $\left\{ \hat{U}_1, \hat{U}_2, \ldots \right\},$ any of which may be invoked with known probabilities $\left\{ p_1, p_2, \ldots \right\}.$ An (unknown) unitary from this collection is selected and can be applied one or more times to a single quantum system, after which a measurement is performed so as to infer the actual unitary used. We consider the case where the \emph{unitary is used once.} This is converted into a standard state discrimination problem~\cite{hellstrom76} by choosing an initial state, described by a density operator $\hat{\rho}_\mathrm{i},$ and applying the given unitary, $\hat{U}_j,$ to yield an output state $\hat{\rho}_{\mathrm{f}\; j} = \hat{U}_j \hat{\rho}_\mathrm{i} \hat{U}_j^\dagger.$ The problem of discrimination between the unitaries reduces to discrimination between the states $\left\{\hat{\rho}_{\mathrm{f}\; 1}, \hat{\rho}_{\mathrm{f}\; 2}, \ldots  \right\}.$ This involves subjecting the system to a POVM with positive operator elements $\left\{ \hat{E}_1, \hat{E}_2, \ldots \right\}$ that satisfy $\sum_j \hat{E_j} = \hat{I}$ and applying an rule for inferring the state  from the measurement outcome (for example, if the measurement yields the outcome associated with $\hat{E_j},$ then the state was $\hat{\rho}_{\mathrm{f}\; j}$). This can be converted to an inference about the unitary used and the task is to find the POVM and the initial density operator which maximizes the probability with which the unitary is correctly inferred. 

The central idea of this work is to use techniques for discrimination of quantum states or unitaries to arrive at and assess quantum algorithms. In the remainder of this article we focus on the Deutsch-Jozsa algorithm.

\section{Application to the Deutsch-Jozsa Algorithm}

The Deutsch-Jozsa problem considers functions $f: \{0,1\}^n \mapsto \{0,1\}$ which are required to be in one of two possible classes : \emph{constant or balanced.} A constant function returns the same output for all possible arguments while a balanced function returns $0$ for exactly half of the possible inputs and $1$ for the other half. The problem is to determine the function class with a minimum number of oracle invocations. A classical algorithm that solves with certainty requires $2^{n-1} +1$ oracle invocations in the worst case~\cite{deutsch92,cleve98}. A quantum algorithm for solving this exists~\cite{deutsch92,cleve98} and, in its modified form~\cite{collins98}, uses the oracle defined on computational basis elements as 
\begin{equation}
	\hat{U}_f \ket{x} := \left( -1\right)^{f(x)}\, \ket{x}.
	\label{eq:oracle}
\end{equation}
The question that we pose is whether it is possible to arrive at quantum algorithms that solve the Deutsch-Jozsa problem purely by considering the possibility of discriminating between the oracle unitaries for balanced and constant functions, given by Eq.~\eqref{eq:oracle}. 

Since the Deutsch-Jozsa problem requires that a determination of function class rather than the actual function used, the discrimination problem is one of discriminating between the two classes of unitary transformations. The can be conveniently recast as a problem of discriminating between two quantum operations, one corresponding to the two constant functions and the other to the set of all balanced functions. Generally either relevant quantum operation can be represented as $\rhoi \mapsto \rhof = \sum_{f\, \textrm{in class}}{p_f} \hat{U}_f \rhoi \hat{U}_f^\dagger$ where $\left\{ p_f | f\, \textrm{in class} \right\}$ are the probabilities with which the various unitaries within each class are applied. The operation for the constant class is independent of these probabilities and performs
\begin{equation}
	\rhoi  \mapsto \rhofconst = \rhoi.
	\label{eq:rhoconst}
\end{equation}
For the balanced class, we consider the case where $p_f$ are identical for all balanced functions. A detailed calculation shows that the operation is
\begin{equation}
  \rhoi  \mapsto \rhofbal = \frac{1}{N-1}\,
	                                  \left( 
	                                       -\rhoi + N \sum_{x=0}^{N-1} \hat{P}_x \rhoi \hat{P}_x
	                                  \right)
	\label{eq:rhobal}
\end{equation}
where $N= 2^n$ and $\hat{P}_x := \ket{x}\bra{x}.$ The problem of discrimination between the two classes of unitaries reduces to that of discrimination between the two density operators, $\rhofconst$ and $\rhofbal.$ Conclusive discrimination between these requires a POVM with two elements $\Econst$ and $\Ebal = \hat{I} - \Econst.$ If the measurement outcome associated with $\Econst$ is obtained, then it will be inferred that the state after the application of the operation is $\rhofconst$ and that hence the unitary is one for a constant function; a similar rule applies for the balanced case. The probability with which a correct inference will be made is $\Prob{\left(\textrm{correct inference} \right)} = p_\mathrm{const} \Trace{\left( \rhofconst \Econst \right)} + p_\mathrm{bal} \Trace{\left( \rhofbal \Ebal \right)}$
%
%
where $p_\mathrm{const}$ and $p_\mathrm{bal} = 1-p_\mathrm{const}$ are the probabilities with which the function is chosen from the constant or balanced classes respectively. The inference will be correct with certainty for arbitrary $p_\mathrm{const}$ if $\Trace{\left( \rhofconst \Econst \right)}= 1$ and $\Trace{\left( \rhofbal \Ebal \right)} =1.$ The requirement that $\Trace{\left( \rhofconst \Econst \right)}= 1$ can be shown, via a series of inferences based on the positivity of both $\rhofconst$ and $\Econst$ and the fact that their eigenvalues are each in the range $[0,1],$ to imply that $\Econst$ is the identity operator on the support of $\rhofconst$ (i.e.\ the subspace orthogonal to the kernel) and is zero elsewhere. Applying this to the analogous operators for the balanced case yields the result that the supports of $\rhofconst$ and $\rhofbal$ are orthogonal or, equivalently 
\begin{equation}
	\rhofconst\rhofbal = \rhofbal\rhofconst =0.
	\label{eq:orthogrho}
\end{equation}
Eqs.~\eqref{eq:rhoconst}-\eqref{eq:orthogrho} imply that 
\begin{align}
   \commut{\rhoi}{\hat{\Lambda}}& = 0 \quad \textrm{and} \label{eq:condone}\\
   N \hat{\Lambda} \rhoi  & = \rhoi^2.  \label{eq:condtwo}
\end{align}
where $\hat{\Lambda} : = \sum_{x=0}^{N-1} \hat{P}_x \rhoi \hat{P}_x $ and satisfies the requirements for a density operator. Denote the orthonormal basis in which the two operators $\rhoi$ and $\hat{\Lambda}$ can be diagonalized simultaneously by $\{ \ket{\phi_j} | j = 1,\ldots N \}$. Thus $\rhoi = \sum_{j=1}^M p_j \ket{\phi_j}\bra{\phi_j}$ where $M$ is the number of non-zero eigenvalues of $\rhoi$ and $\hat{\Lambda} = \sum_{j=1}^N \lambda_j \ket{\phi_j}\bra{\phi_j}$. Suppose that $p_1\neq 0.$ Then Eqs.~\eqref{eq:condone} and~\eqref{eq:condtwo} can be shown to imply that
\begin{equation}
  \frac{p_1}{N} =  p_1 \sum_{x} \lvert \phi_1(x) \rvert^4
                   + \sum_{k \neq 1}^M p_k \sum_{x} \lvert \phi_k(x) \rvert^2 \lvert \phi_1(x) \rvert^2.
  \label{eq:mainconstraintonp}
\end{equation} 
where $\phi_1(x)= \innerprod{x}{\phi_1}.$ The normalized state $\ket{\phi_1}$ satisfies $\sum_{x} \lvert \phi_1(x) \rvert^2 =1$ and a Lagrange multiplier technique demonstrates that, subject to this constraint, $\sum_{x} \lvert \phi_1(x) \rvert^4 \geqslant 1/N$ with equality attained if and only if $\lvert \phi_1(x) \rvert^2 = 1/N$ for all $x=0,\ldots N-1.$ Thus Eq.~\eqref{eq:mainconstraintonp} implies
\begin{equation}
  \frac{p_1}{N} \geqslant \frac{p_1}{N}
                   + \sum_{k \neq 1}^M p_k \sum_{x} \lvert \phi_k(x) \rvert^2 \lvert \phi_1(x) \rvert^2.
  \label{eq:mainconstraintonptwo}
\end{equation} 
with equality if and only if $\lvert \phi_1(x) \rvert^2 = 1/N$ for all $x=0,\ldots N-1.$ Since the second term on the right hand side of Eq.~\eqref{eq:mainconstraintonptwo} is non-negative, the only possibility is that the equality holds and that $\lvert \phi_1(x) \rvert^2 = 1/N$ for all $x=0,\ldots N-1.$ This implies that $p_k = 0$ for $k\neq 1.$ 

Thus the only possible initial states which discriminates conclusively and correctly with certainty between the two classes of quantum operations for the Deutsch-Jozsa problem are $\rhoi = \ket{\phi}\bra{\phi}$ where 
\begin{equation}
	\ket{\phi} = \frac{1}{\sqrt{N}}\, \sum_{x=0}^{N-1} e^{i\theta_x} \ket{x}
\end{equation}
with $\theta_x$ arbitrary real phases. The corresponding POVM elements are $\Econst = \ket{\phi}\bra{\phi}$ and $\Ebal = \hat{I} - \Econst.$ It is easily verified that this algorithm discriminates with certainty between balanced and constant functions regardless of \emph{a priori} probabilities.

\section{Conclusion}

We have shown that the techniques of unitary discrimination can be applied to yield an exhaustive collection of algorithms which solve the Deutsch-Jozsa problem with certainty. This suggests that it may be fruitful to investigate unitary discrimination in the context of other algorithms or whenever the set of input states is restricted (such as thermal equilibrium states in NMR).




\begin{theacknowledgments}
 The author is grateful for support from the Mesa State College Faculty Professional Development Fund.
\end{theacknowledgments}



\bibliographystyle{aipproc}   


\IfFileExists{\jobname.bbl}{}
 {\typeout{}
  \typeout{******************************************}
  \typeout{** Please run "bibtex \jobname" to optain}
  \typeout{** the bibliography and then re-run LaTeX}
  \typeout{** twice to fix the references!}
  \typeout{******************************************}
  \typeout{}
 }

\end{document}